\documentclass[conference]{IEEEtran}
\usepackage{graphicx}
\usepackage{amsmath}
\usepackage{color}
\usepackage{cite}

\begin{document}
\title{A Review on Quantum Computing: \\ Qubits, Cryogenic Electronics \\ and Cryogenic MOSFET Physics}

\author{\IEEEauthorblockN{Farzan Jazaeri$ ^{a} $, Arnout Beckers$ ^{a} $, Armin Tajalli$ ^{b} $, and Jean-Michel Sallese$ ^{a} $}
	\IEEEauthorblockA{a) Ecole Polytechnique F\'ed\'erale de Lausanne (EPFL), Switzerland.\\b) Electrical and Computer Engineering Department, University of Utah, USA.\\ E-mail: farzan.jazaeri@epfl.ch, arnout.beckers@epfl.ch}}

\IEEEspecialpapernotice{(\textbf{Invited Paper})}

\maketitle

\begin{abstract}
Quantum computing (QC) has already entered the industrial landscape and several multinational corporations have initiated their own research efforts. So far, many of these efforts have been focusing on superconducting qubits, whose industrial progress is currently way ahead of all other qubit implementations. This paper briefly reviews the progress made on the silicon-based QC platform, which is highly promising to meet the scale-up challenges by leveraging the semiconductor industry. We look at different types of qubits, the advantages of silicon, and techniques for qubit manipulation in the solid state. Finally, we discuss the possibility of co-integrating silicon qubits with FET-based, cooled front-end electronics, and review the device physics of MOSFETs at deep cryogenic temperatures. 
\end{abstract}
\vspace{-0.1cm}
 
\IEEEpeerreviewmaketitle
\section{Introduction}
Quantum computing is attracting more and more the interest of industrial actors, not only “broad-interest” corporations like Google \cite{1} or Microsoft\,\cite{2}, but also companies more traditionally linked to the area of nanoelectronics and nanotechnology (e.g., IBM~\cite{3}, Intel~\cite{intel}, and Lockheed Martin~\cite{4}).

The field of quantum computing started in the early 1980s. The need for quantum computers (QCs) to simulate quantum physics efficiently, was first foreseen by Richard Feynman~\cite{Feynman1982}. His focus at that time was primarily on the application of QCs to simulate physics, notably quantum mechanics, by exploiting the intrinsic parallelism present in quantum systems. In 1985, David Deutsch proposed to model a universal quantum computer in his seminal paper, ``\emph{Quantum theory, the Church-Turing principle and the universal quantum computer}''~\cite{deutsch1985quantum}. The years following this publication saw several developments in quantum algorithms and quantum-computing theory, particularly by Simon and Vazirani. Interest in the QC field increased exponentially after Peter Shor discovered his factoring algorithm and Lov Grover his search algorithm~\cite{Shor23,365700, Grover2, PhysRevLe3}. Consequently, the breadth of applications of QCs expanded from Richard Feynman's initial proposal of quantum chemistry/physics, to other potential applications that speak to a wider audience: information security, optimization, machine learning, etc. Later on, in 1999, the world's first quantum computer was made out of superconductors by a Canadian company called D-Wave Systems~\cite{DWave, 5}. A 28-qubit quantum computer was demonstrated in 2007, followed by 128 qubits in 2010, 512 qubits in 2013, and 2000 qubits in 2018. However, there is still a heavy debate going on about the actual \textquoteleft quantumness\textquoteright \, of the D-wave computers~\cite{debate, dbate2}. The approach followed by D-Wave Systems can be viewed as similar to early computers in the sense that their quantum computer is non-universal, meaning that it can only perform very specific tasks~\cite{debate}. That is why some people are convinced that a \textquoteleft universal\textquoteright\, quantum computer is yet to be realized. Very recently, IBM proclaimed their company goal to build commercially available, universal quantum computing systems. Their first IBM-Q systems and services are now delivered via the cloud. In addition to Lockheed Martin and IBM, other companies such as Google, Intel, and Microsoft have joined the race for a universal  quantum computer. They all have launched recently major research efforts on quantum computing, funding laboratories all over the world known for their early developments in QC. Besides private companies, the European Commission has recently announced the billion-dollar Quantum Technologies Flagship~\cite{flagship}. Smaller European projects are ongoing as well, e.g., the MOS-Quito Project (MOS-based Quantum Information Technology)~\cite{mosquito}, which was set up after the birth of the CMOS compatible qubit in France~\cite{maurand}. Many other physical realizations of quantum bits (qubits) have been proposed and investigated at the level of basic research laboratories. Solid-state implementations have gained increasing attention in recent years owing to their potential for scaling to larger numbers of qubits. Solid-state qubits appear to be the overwhelming choice for industrial and commercial R\&D efforts. To the group of solid-state qubits belong superconducting Josephson junctions~\cite{Devoret1169} and spin qubits in silicon~\cite{article3,pla2012single}. The compatibility of silicon qubits with CMOS foundries is a great asset. To increase the number of qubits in commercial quantum computers, the number of interconnections to the measurements equipment at room temperature will become intractable very soon. Each qubit needs to be individually addressed, requiring a single connection to the outside world. This lowers the degree to which the sub-Kelvin environment of the qubits can be isolated. Thermal noise seeping into the fridge is nefast for the quantum computations. The qubits reside at the bottom of the dilution refrigerator. The interconnections pass from the $\approx$\,10-mK environment to the 4.2-K stage of the dilution fridge all the way to outside. Since thermal noise can never be completely avoided, the current view is that quantum error correction is necessary, introducing even more qubits for redundancy. Several redundant physical qubits have to be used to encode one logical qubit. Furthermore, the wiring capacitances accumulate when the system scales, which increases the latency. This poses problems to running more sophisticated quantum algorithms that require immediate action after a qubit is read~\cite{bronn}. To face these scale-up challenges, the use of CMOS front-end electronics that can operate at deep cryogenic temperatures ($<$\,10\,K) has been proposed by many research groups all over the world~\cite{tracy_single_2016,ekanayake,charbon,vandersypen2017interfacing,hornibrook,degenhardt,veldhorst2017silicon}. Recently, a CMOS prototype IC with a pulse generator to talk to qubits has been deployed in the 4-K stage of the Google Bristlecone quantum computer~\cite{bardin}. By sharing the same CMOS technology, qubits and peripheral electronics could lie even on the same chip. To fully profit from the most established industrial technology, the ultimate goal would be to realize so-called \textquoteleft quantum-integrated circuits\textquoteright \, or QICs. Many challenges remain to accomplish this, ranging from qubits to MOSFET models to system architectures. 

In this review, we discuss when (not) to use quantum computers, the physics of qubits, basic quantum logic gates, different physical realizations of qubits, and the development of peripheral low-temperature electronics for an efficient interface between qubits and external (room-temperature) circuitry. Finally, we explain the MOSFET physics at deep cryogenic temperatures that furthers our understanding for model verification and validation for deep cryogenic IC design. 

\section{When (not) to Use Quantum Computers}
Quantum computers will only give a speed-up over conventional computers for specific problems, typically \textquoteleft very hard\textquoteright \, problems. The quantum computer works by executing the same quantum algorithm several times over again. The most likely result after these runs is the solution. The time it takes to run the quantum computer several times can still be exponentially faster to arrive at the result of a very hard problem than a conventional computer would need in a single run working on the same problem. Many problems are characterized by this \textquoteleft very hard\textquoteright\, exponential growth in complexity. These problems are intractable today on supercomputing clusters in a reasonable amount of time. We can thus benefit from quantum computers in optimization problems, machine learning, sampling of large data sets, forecasting etc. Another example is quantum chemistry (e.g., simulating proteins for new medicine) that is currently running at the limit of classical computers. To actually simulate what is going on in a simple molecule, every electron-electron interaction needs to be taken into account. Next, Shor's algorithm is the most famous example of a computation problem that requires a quantum computer to solve. To factorize a number of $N$ into its prime numbers, a classical computer would take, in some cases, more than the age of the universe to produce a result. It is, therefore, used as a basic encryption tool in information security all over the world. A quantum computer could solve this problem much faster. Lastly, chip manufacturers tend to go to great lengths to suppress quantum effects. It seems only natural to attempt to harness these quantum effects for the benefit of computing. 

\section{Qubits and Quantum Logic Gates}
\subsection{The Qubit}
\begin{figure}[b]
	\centering
	\includegraphics[width=0.3\textwidth]{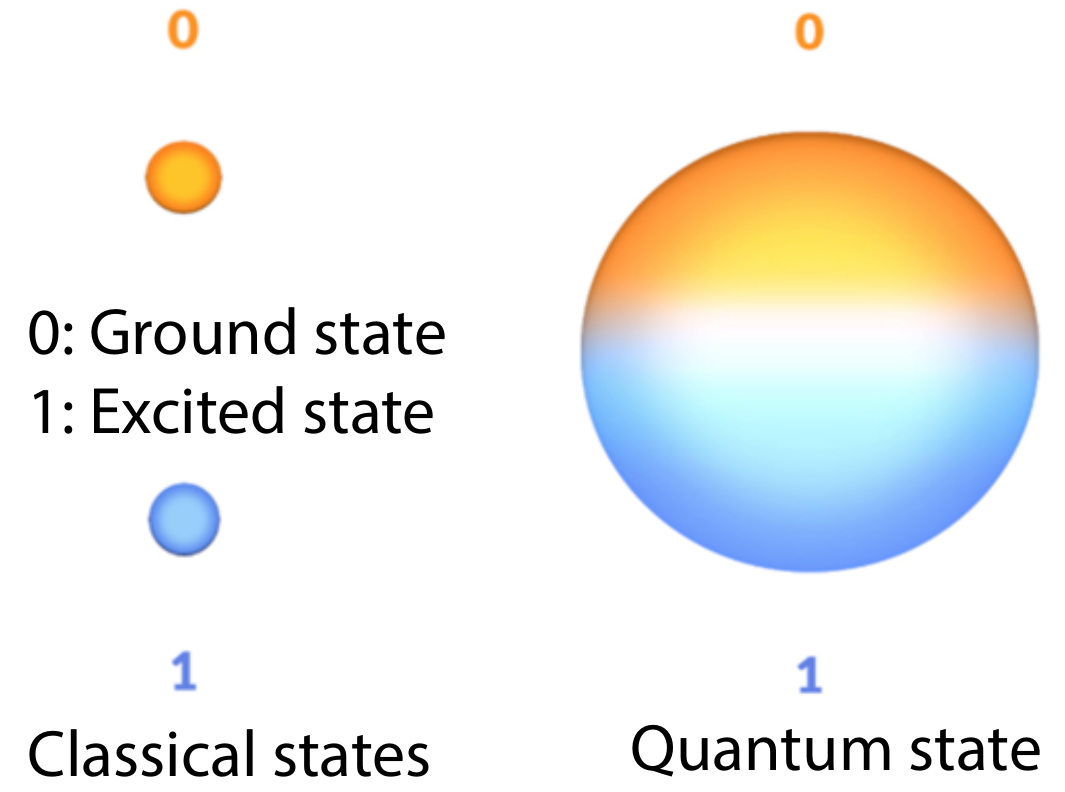}
	\caption{The basic unit of information in a QC is the quantum bit (qubit) which can be in any linear combination of ground and excited sates.}
	\label{Fig1}
\end{figure}

The qubit (or quantum bit) is the basic container of information in a QC, replacing the bit in a conventional computer. The qubit can be in both ground and excited states at the same time (see Fig. \ref{Fig1}). The two logical states of each qubit must be mapped onto the eigenstates of some suitable physical system. The most straightforward example is the spin. A spin qubit relies on a spin degree of freedom of either electronic or nuclear nature, which can hold a bit of quantum information for very long times. Note that there are many other examples of qubits: two different polarizations of a photon, two energy states of an electron orbiting a single atom, etc. The quantum computer is fundamentally different than a classical computer due to two distinct properties of qubits. The first property is \textquoteleft quantum superposition\textquoteright \, or the linear combination of possible configurations. The second one is \textquoteleft quantum entanglement\textquoteright. 
\subsection{Quantum Superposition}
\begin{figure}[t]
	\centering
	\includegraphics[width=0.45\textwidth]{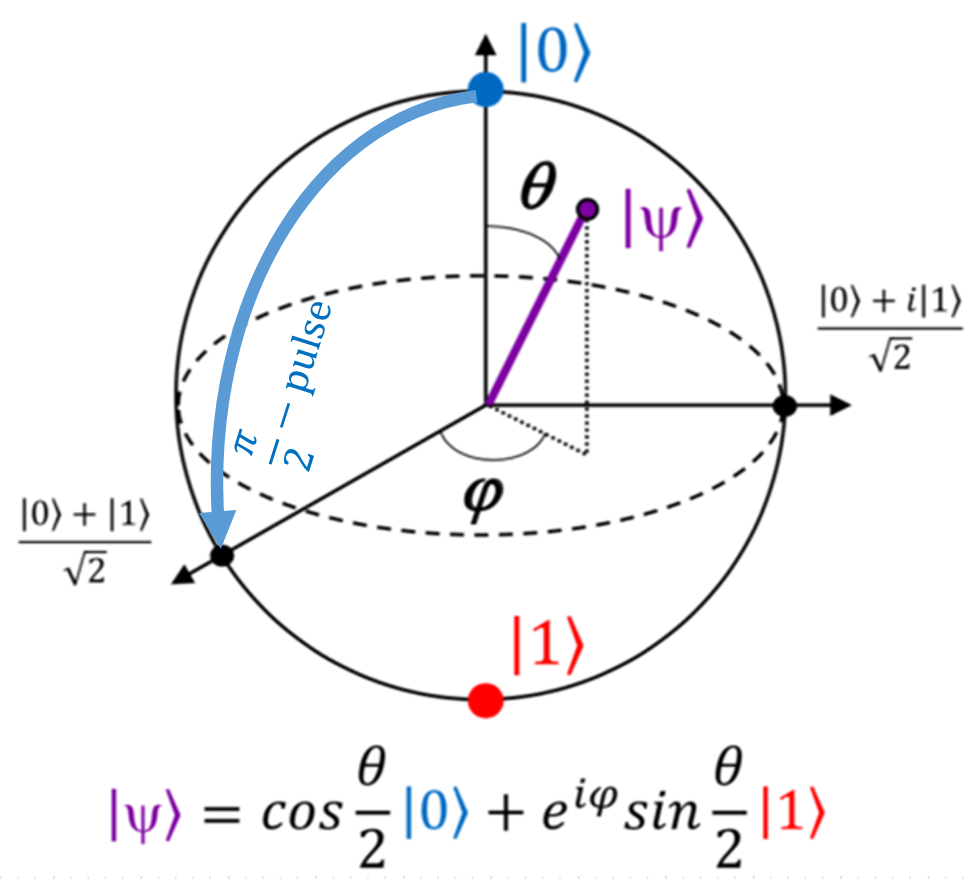}
	\vspace{-0.2cm}
	\caption{The Bloch sphere provides a useful means of visualizing the state of a single qubit and operations on it. Any point on this sphere represents a linear combination of the 0 and 1 states with complex coefficients. A $\pi/2$-pulse \textquoteleft rotates\textquoteright \, a qubit from the 0-state to a superposition state.}
	\label{Fig2}
\end{figure}

Consider a system with two basis states, call them $ |0\big> $ and $ |1\big> $. A classical bit of data can be represented by a single atom that is in one of the two states denoted by $ |0\big> $ and $ |1\big> $ (left of Fig.\,\ref{Fig1}).  By contrast, a quantum state of a qubit is in a continuous state between ``0'' and ``1'' until the qubit is measured. The outcome can only be ``0'' or ``1''. Therefore, a qubit is a continuous object and its quantum state is given by $|\psi\big>=\alpha_1|0\big>+\alpha_2|1\big>$, where $\alpha_1$ and $\alpha_2$ are complex amplitudes. If we measure this in the computational basis, we obtain the $ |0\big> $ state with probability $ |\alpha_1|^2$ or the $ |1\big> $ state with probability $|\alpha_2|^2$, where  $|\alpha_1|^2+|\alpha_2|^2=1 $. If one qubit can be in the superposition of two classical states, two qubits can be in a superposition of four, and $n$ qubits can be in a superposition of $ 2^n $: $|0\big> $, $ |1\big> $, $ |2\big> $,...,$ |n-1\big> $. Therefore, a quantum register of $n$ qubits is given by $|\psi\big>=\alpha_0|0\big>+\alpha_1|1\big>+...+\alpha_{2^n-1}|2^n-1\big>$, where $\sum_{j=0}^{2^n-1}|\alpha_j|^2=1$. Figure \ref{Fig2} shows the Bloch sphere. The Bloch sphere provides a useful means of visualizing the state of a single qubit and often serves as an excellent testbed for ideas about quantum computation and quantum information~\cite{chuang}. The angles $ \theta $ and $ \varphi $ can be interpreted as the polar and azimuth angles of points on the sphere, respectively. There are an infinite number of these points on the unit sphere. It is worth emphasizing that we cannot ``see'' a superposition itself, but only classical states. It is not determined in advance what the outcome of the measurement will be. The only thing we can say before the measurement is that we will observe state $|j\big>$ with probability $ |\alpha_j|^2 $. 
\subsection{Quantum Entanglement} 
The second property that distinguishes QCs from conventional computers is entanglement. If two qubits are \textquoteleft entangled\textquoteright \, there is a correlation between these two qubits. If one qubit is in one particular state, the other one \emph{has to be} in another particular state. If two electrons become entangled, their spin states are correlated such that if one of the electrons has a spin-up, then the other one has a spin-down after measurement. This property was pointed by Albert Einstein in 1935~\cite{einstein}. The creation and manipulation of entangled states plays a central role in quantum information processing.  

\subsection{How to Interact with a Qubit?}
For example, an electron in an atom can be in either the ground state or any excited state. By shining light on the atom, with appropriate energy and for an appropriate length of time, it is possible to move the electron from the ground to excited state or vice versa. More interestingly, by reducing the time we shine the light, an electron initially in the state $ |0\big> $ can be moved halfway between $ |0\big> $ and $ |1\big> $. 

Similarly to the previous example, the qubits in a quantum computer are first initialized, say all spin-up (zero state on the Bloch sphere in  Fig.\,\ref{Fig2}) by applying a large, static magnetic field of around one Tesla. An electromagnetic pulse is then applied to each qubit individually to bring the qubit into any possible state on the Bloch sphere. The actual qubit state depends on the pulse amplitude and time. For instance, a superposition state can be created when using a so-called $\pi/2$-pulse, which \textquoteleft rotates\textquoteright \, the qubit from the top of the Bloch sphere to the equator, as shown in Fig.\,\ref{Fig2}. The timing of a $\pi/2$-pulse is given by $t=\pi/(2\omega)$, where $\omega$ is the angular frequency of the electromagnetic driving field~\cite{lebellac}. Similarly, a $\pi$-pulse would flip the qubit from the zero to the one state in Fig.\,\ref{Fig2}. For silicon qubits, $\omega$ is typically in the   
order of hundreds of MHz to tens of GHz. These type of pulses can be created by oscillators integrated in advanced CMOS technologies, as discussed in Sec.\,\ref{sec:elec}. The superposed qubits then still need to be coupled (entangled). The quantum information is in the fragile entanglements. After time evolution, the quantum states are read-out and the quantum algorithm is repeated. 

To sum up this subsection, there are two main things we can do with a qubit: (i) we can measure it as 0 or 1 with a certain probability, or, (ii) we can apply some operation to it, which corresponds to rotating the quantum state to a different position on the Bloch sphere. Rotations on the Bloch sphere are implemented in a quantum computer by so-called \textquoteleft quantum logic gates\textquoteright \, which are discussed next. 
\subsection{Quantum Logic Gates}
To build a universal quantum computer, a set of universal quantum logic gates is required, similar to the ones present in classical computers~\cite{deutsch1985quantum}. A unitary operator that acts on a small number of qubits is often called a \textquoteleft gate\textquoteright\,, in analogy to classical logic gates like AND, OR, and NOT. In what follows, we discuss three of the most important quantum logic gates. 

\subsubsection{NOT Gate}
We start with a simple quantum NOT gate, also called bitflip gate, or $X$. The output of this gate is $ c_1|0\big>+c_0|1\big> $  when the input is $ c_0|0\big>+c_1|1\big> $. In matrix form, this is
\begin{equation}
X=
\begin{pmatrix} 
0 & 1 \\
1 & 0 
\end{pmatrix}
\quad
\end{equation}\label{Eq2a}
\subsubsection{Controlled-NOT Gate}
An example of a two-qubit gate is the controlled-not gate or CNOT. It negates the second bit of its input if the first bit is $ |1\big> $, and does nothing if the first bit is 0. Therefore, CNOT is a quantum gate which is different than NOT and the output depends on the first input. The first qubit is called the control qubit, the second the target qubit. In matrix form, this can be expressed as:
\begin{equation}
CNOT=\dfrac{1}{\sqrt{2}}\quad
\begin{pmatrix} 
1 & 0&0&0 \\
0 & 1&0&0 \\
0 & 0&0&1 \\
0 & 0&1&0 \\
\end{pmatrix}
\quad
\end{equation}\label{Eq3-a}
\subsubsection{Hadamard Gate}
Another important quantum gate is the Hadamard gate which is used to create the quantum superposition ($\pi/2$-pulse), specified by $H|0\big>=\frac{1}{\sqrt{2}}|0\big>+\frac{1}{\sqrt{2}} |1\big>$
when starting from the zero state. 
The Hadamard gate outputs $|0\big> $ or $ |1\big>  $ with equal probability. As a unitary matrix, this is 
\begin{equation}
H=\dfrac{1}{\sqrt{2}}\quad
\begin{pmatrix} 
1 & 1 \\
1 & -1 
\end{pmatrix}
\quad
\end{equation}
There are other quantum gates in use, such as the phaseflip, phaseshift, and Toffoli gate among others. For a more detailed treatment of quantum gates, see e.g., Chuang and Nielsen~\cite{chuang}. 
\section{Qubit Implementations}
The qubits in a quantum computer should meet the following five traditional requirements called the DiVincenzo criteria, after the theoretical physicist David P. DiVincenzo~\cite{DiVincenzo}: 
\begin{enumerate}
	\item A scalable physical system,
	\item Ability to initialize the state of the qubits,
	\item Coherence time longer than the gate operation time,
	\item A universal set of quantum gates,
	\item A qubit-specific measurement capability.
\end{enumerate}
Over the years, researchers have come up with several physical realizations of qubits that each fulfill these criteria to a certain extent. Liquid-state nuclear-magnetic-resonance (NMR) \cite{vandersypen2005nmr} and ion traps \cite{article2} were amongst the earliest studied. Superconducting and semiconductor qubits are becoming more relevant for scalability. Each qubit type has its own advantages and industrial players have placed bets on their favorites.  

\subsection{Nuclear Magnetic Resonance Qubits}
Nuclear Magnetic Resonance (NMR) provides a way to build QCs by using the spin of atomic nuclei\cite{Cory1634}. Shor's factoring algorithm has been realized using this technique by Vandersypen and Chuang~\cite{vandersypen2001experimental}. NMR systems have fairly long relaxation times, and thus decoherence is not a major problem. Even though NMR largely satisfies the DiVincenzo criteria reaching a high degree of qubit control and permitting hundreds of operations in sequence, there is a quite strong limitation on the size that NMR quantum computers can attain.

\subsection{Ion Traps}
Ion traps offer great possibilities for building quantum computers \cite{article2}. Ion trap quantum computing operates on a qubit register formed by a linear string of ions confined in a high electromagnetic field. Ions are excellent quantum memories and long coherence times have been demonstrated~\cite{article2, 5088164}. In addition, scaling the number of qubits in an ion trap quantum processor is in principle feasible~\cite{Monroe1164}. 

\subsection{Superconducting Qubits}
Scientists have built artificial atoms made out of superconducting devices~\cite{Devoret1169}, e.g., a Josephson junction, coupled to a microwave resonator for control and read-out~\cite{3}. This is the road chosen by e.g., Google and IBM. A normal LC circuit made from non-superconducting metals is not a good qubit because the energy spacing of the quantum harmonic oscillator is the same over the energy scale ($E=\hbar \omega_o$, where $\omega_o=\sqrt{LC}$ is the resonance frequency). The Josephson inductance introduces a nonlinearity which increases the energy spacing such that the ground state and first excited state of the oscillator become separate from the other excited states. This allows to selectively excite a two-state system and thus use it as a qubit.

\subsection{Nitrogen Vacancies in Diamond}
Nitrogen vacancies (NV centers) in diamond have the advantage over other qubits in that they are easily coupled to photons. This can be exploited to accomplish efficient quantum networks~\cite{childress_hanson_2013}. It is not unlikely that different types of qubits will complement each other in different systems and applications. However, NV centers are defects in a diamond crystal which are harder to control during fabrication than spin qubits in silicon, including the CMOS compatible qubits. 

\subsection{\label{sec:sil}Silicon Qubits and CMOS Compatible Qubits}
\begin{figure}[t!]
	\includegraphics[width=0.5\textwidth]{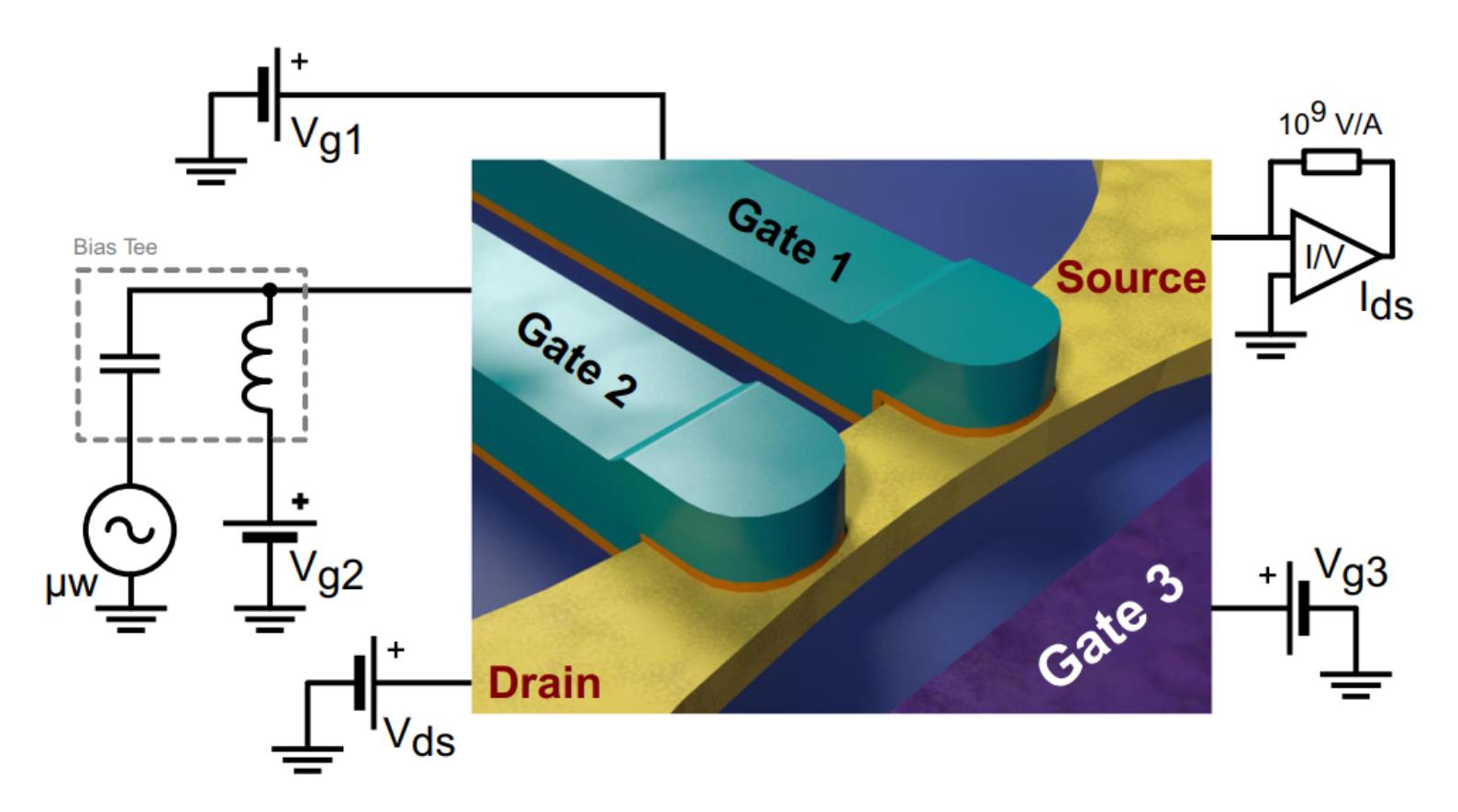}
	\vspace{-0.5cm}
	\caption{Schematic of a double quantum dot system. From Corna \emph{et al.}~\cite{article23}.}
	\label{fig4}
\end{figure}

Semiconductor qubits were initially developed in III-V materials, typically GaAs~\cite{bluhm2011dephasing}. However, a variety of spin qubits in silicon have now been demonstrated in academic research laboratories that have longer spin coherence times, up to several milliseconds~\cite{schreiber2014quantum}. This is mainly thanks to the possibility of isotopically purifying silicon from excess nuclear spins. Nuclear spins in the host material can decohere the quantum state of a qubit. In silicon, two routes for spin qubit fabrication are available: qubits that rely on (i) the spins of electrons in electrostatically defined quantum dots (QDs), and (ii) the spins of electrons or nuclei of impurity atoms implanted in the silicon host. Apart from its outstanding properties as a host material for qubits, silicon has the advantage that the processes to make these quantum devices are largely in place today, requiring relatively few modifications to existing silicon fabs. Silicon is also the best understood material in terms of defects by years of industrial experience. The University of New South Wales in Sydney proposed the first qubits in silicon~\cite{pla2012single}. A few years later, CEA-L\'eti in France proposed the first CMOS compatible qubit in a fully-depleted SOI technology~\cite{maurand,hutin}. Others are following the initiative to use silicon. For instance, the Hughes Research Laboratory in the United States has recently opened a research line on silicon-based quantum computing. Furthermore, the nanoelectronics expertise center IMEC in Belgium is collaborating with several partners, including CEA-L\'eti, to scale up the fabrication of quantum devices in its state-of-the-art cleanrooms~\cite{imecjoins}. As shown in Fig.\,\ref{fig4}, the main idea of a CMOS compatible qubit is to use a MOS gate to electrostatically confine electrons in a corner quantum dot~\cite{betz}. Only one half of a gate is required for one dot. A single electron can be trapped in the corner by repelling electric fields imposed from all sides. Normally, we would interact with a spin qubit by drawing a metal line close to the qubit and pushing an oscillating current through the wire which would produce an oscillating magnetic field that can interact with the electron spin. However, the use of a metallic wire for each qubit is not very scalable. It is possible to remove the wire and use only a MOS gate to control the spins. This technique is known in literature as Electric Dipole Spin Resonance (EDSR). However, using only a MOS gate, we do not have an oscillating magnetic field and the electric field on the gate interacts only with the position of the electron, not directly with its spin. A requirement to achieve coupling to the electron's spin is that a material with a strong spin-orbit coupling is used. To achieve a high spin-orbit coupling in silicon, the focus has recently shifted from electron spin qubits to hole spin qubits~\cite{crippa}.  

\begin{figure}[t]
	\includegraphics[width=0.5\textwidth]{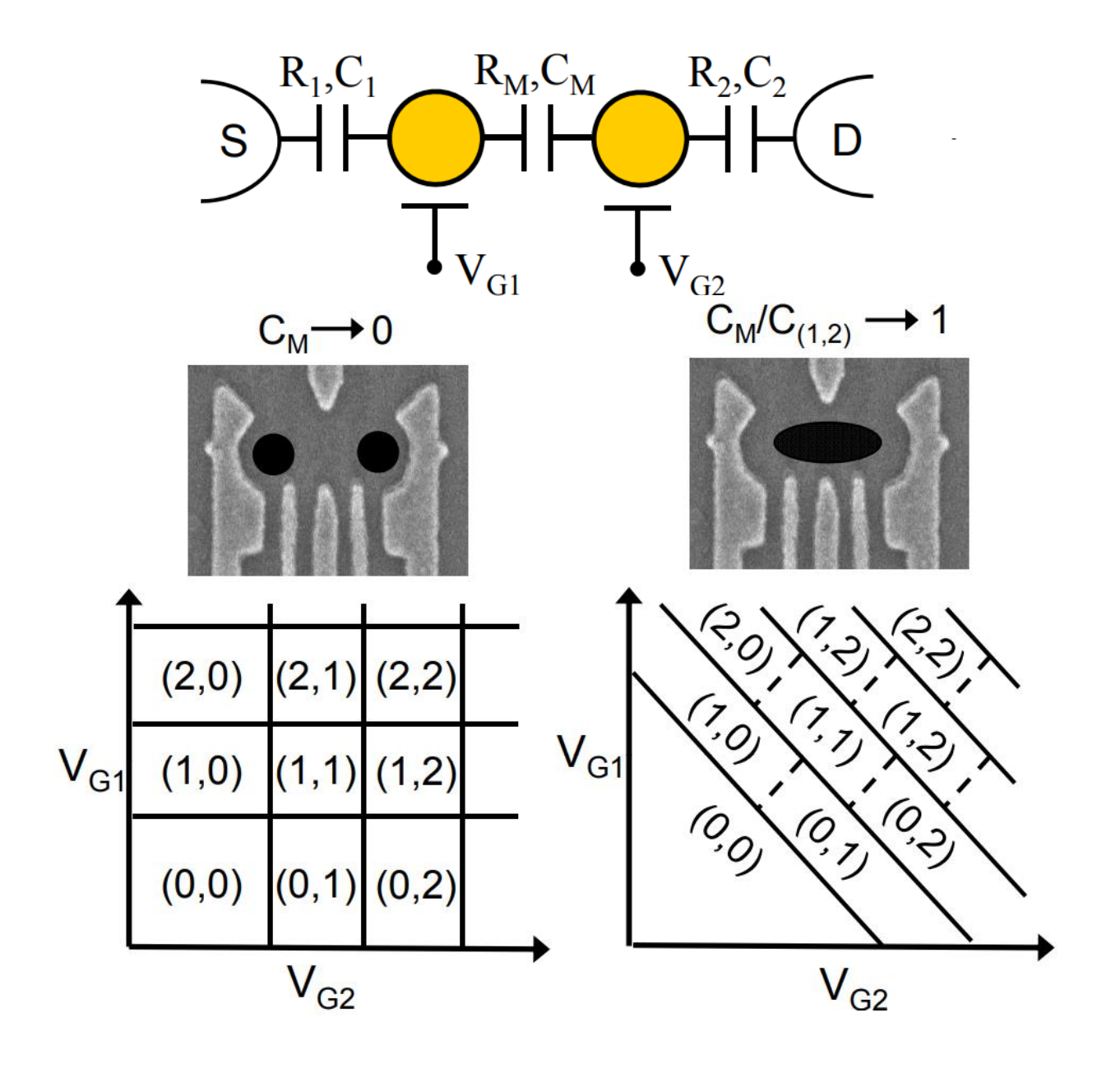}
	\vspace{-1cm}
	\caption{Network of tunnel resistors and capacitors representing two quantum dots coupled in series (top). Charge stability diagrams for (left) uncoupled and (right) coupled double dots, depicting the equilibrium electron numbers 		(N1,N2) in dot 1 and 2 respectively. Adapted from \cite{RevModPhys.75.1}}
	\label{fig3}
\end{figure}

Let's first have a look at the double quantum dot system as shown on the top of Fig. \ref{fig3}. Double dots are useful in QC experiments because one dot can be used for manipulation and the other for measurement, or both can be used to investigate entanglement.
It consists of two electrodes known as the drain and the source, we have two quantum dots between the source and drain contacts. The electrical potential of the quantum dots can be tuned by  electrodes, known as the gates, which are capacitively coupled to the islands. In double quantum dots, electrons can be transferred from one quantum dot to the other by modifying the electrostatic potential landscape using gate voltages. Depending on the coupling capacitance between the quantum dots, we have to different situations. On the left side of Fig. \ref{fig3}, no coupling exists between the two quantum dots. Different rectangular regions in stability diagrams correspond to different numbers of electrons. On the right side of Fig. \ref{fig3}, when the two quantum dots are coupled, the rectangular regions are symmetric along the line for $ V_{G1}= V_{G2}$. 

During the transport through a dot, the number of electrons in the dot is varied one by one which is a single electron tunneling. This means that the N electron ground state can only be constructed by adding (subtracting) one electron to (from) N-1 (N+1) electron ground state. In such quantum system, a blocking state might occur due to the coulomb blockade and therefore no accessible energy levels are within tunneling range of an electron on the source contact. 
Variations in the charge of quantum dots due to the tunneling transitions cause a small shift in the resonance frequency which can be detected in the amplitude and phase of the reflected signal in RF reflectometry technique  RF reflectometry technique with resonant frequency of few hundred MHz is used to control and measure qubits if we are not dealing with the spin state of the electrons.  

For a given dot structure, a charge stability diagram, or a ``honeycomb diagram'', can be formed, telling us at which gate voltages what electron occupancy is favored. To construct this diagram, we first model our double or triple dot as a system of capacitors (see Fig. \ref{fig3}). In a perfectly uncoupled double dot, we would expect a pattern of rectangles to be formed in our stability plot. In a perfectly coupled system, we would expect rectangles again, however, now they would be symmetric about the $ V_{G1} = V_{G2} $. 
Here, we can flip the orientation of the magnetic moments through the use of electromagnetic radiation at resonant frequencies. The valley eigenstates are separated by a magnetic field and the Pauli blockade regime can be achieved. Therefore, QD1 acts as an effective “spin filter” regulating the current flow induced by EDSR in QD2. Pauli blockade has been utilized to implement spin-to-charge conversion for reading spin states of electrons in double QDs.
\begin{figure*}[t!]
	\includegraphics[width=\textwidth]{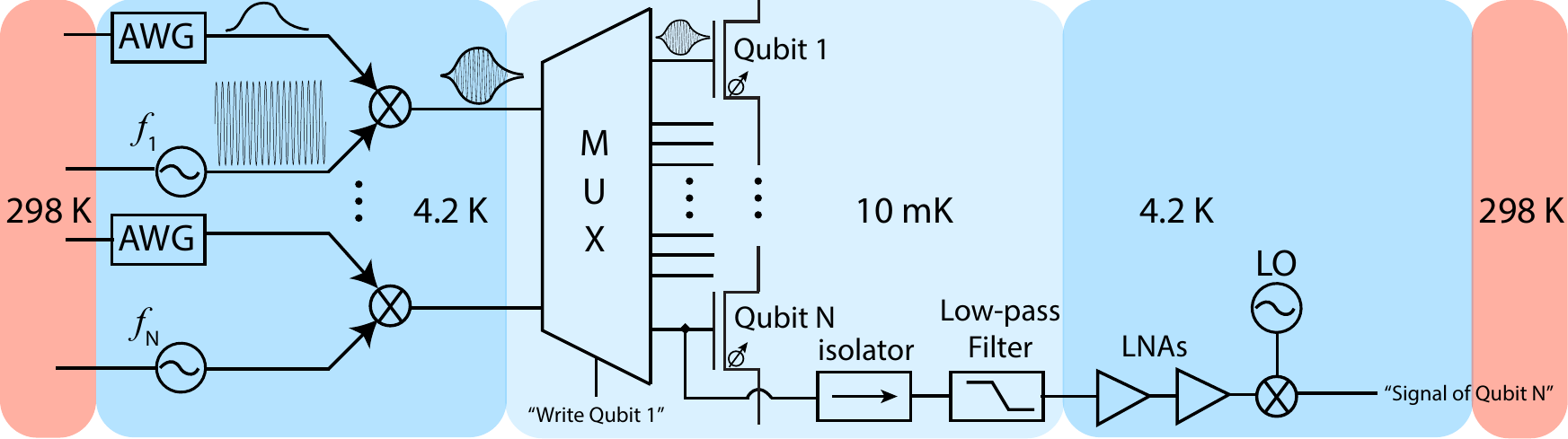}
	\caption{Simplified schematic showing the front-end electronic circuits required for controlling (\textquoteleft writing\textquoteright) and measuring (\textquoteleft reading\textquoteright) spin qubits.}
	\label{fig:elec}
\end{figure*}
\section{\label{sec:elec}Cryo-Cooled Front-end Electronics for Qubits}
In commercial QCs, the equipment that is needed for qubit control and read-out (i.e., shaping electrical pulses, amplification, etc.) resides mainly at room temperature. As mentioned in the Introduction, this brings about many challenges. These challenges can be at least partially alleviated by integrating the same functionality with CMOS circuits in silicon chips, and implementing them in the 4-Kelvin stage of the dilution refrigerator of the quantum computer. This can reduce latency and increase overall system scalability. If we would extrapolate this practice, ultimately one would end up with so-called quantum integrated circuits (QICs)  where solid-state qubits (most likely silicon qubits) and peripheral electronics are monolothically co-integrated in the same microsystem. With successful future engineering of qubit devices, it might become possible to increase the operating temperature of the qubits from $\approx$\,10\,mK to 4.2\,K and then operate the whole system at 4.2\,K. At 4.2\,K, the cooling power is several orders of magnitude higher than at 10\,mK and more functionality can thus be implemented at 4.2\,K. Figure~\ref{fig:elec} shows a simplified schematic that illustrates some of the required peripheral electronics to read and write spin qubits. The \textquoteleft Write\textquoteright\, operation of spin qubits (aka controlling, manipulating, or rotating) is shown on the left. The \textquoteleft Read\textquoteright\, operation (aka measuring) is shown on the right. To write a qubit, an RF pulse needs to be shaped with a well-controlled timing and amplitude and sent to the qubit. This can be done with an arbitrary wave-form generator (AWG), an RF oscillator, and a mixer. The AWG can be designed in CMOS as a digital-to-analog converter, the oscillator as a digital ring oscillator~\cite{ringoscil} or a voltage-controlled oscillator, and the mixer as a single nonlinear element, e.g., a MOSFET. Multiplexers allow to address multiple qubits at once. Frequency-division and time-division multiple-access techniques are also being investigated for this purpose. To read a qubit, typically RF reflectometry is used on the MOS gate, as briefly mentioned in Sec.\,\ref{sec:sil}. A read-out pulse is sent to a qubit, which picks up a frequency or phase shift depending on the state of the qubit. The RF read-out signal travels through the isolator, low-pass filter, and amplifier chain, before it reaches the down-converter for further processing. Other circuits such as current and voltage references~\cite{voltagereference} will also be necessary to realize QICs. Note that Fig.\,\ref{fig:elec} is a simplified schematic and that many other architectures are under research. For instance, recently a structure has been proposed that resembles a one-transistor-one-capacitor dynamic random-access memory technology~\cite{schaal2019cmos}. The qubit is placed on the spot of the capacitor and the MOSFET is used to read-out the state of the qubit. From the previous discussion, it has become clear that predicting the performance and power consumption of MOSFETs at deep cryogenic temperatures is essential. Digital, analog and RF models are required to meet trade-offs in the design phase.
\begin{figure*}[t!]
	\centering
	\includegraphics[width=0.6\textwidth]{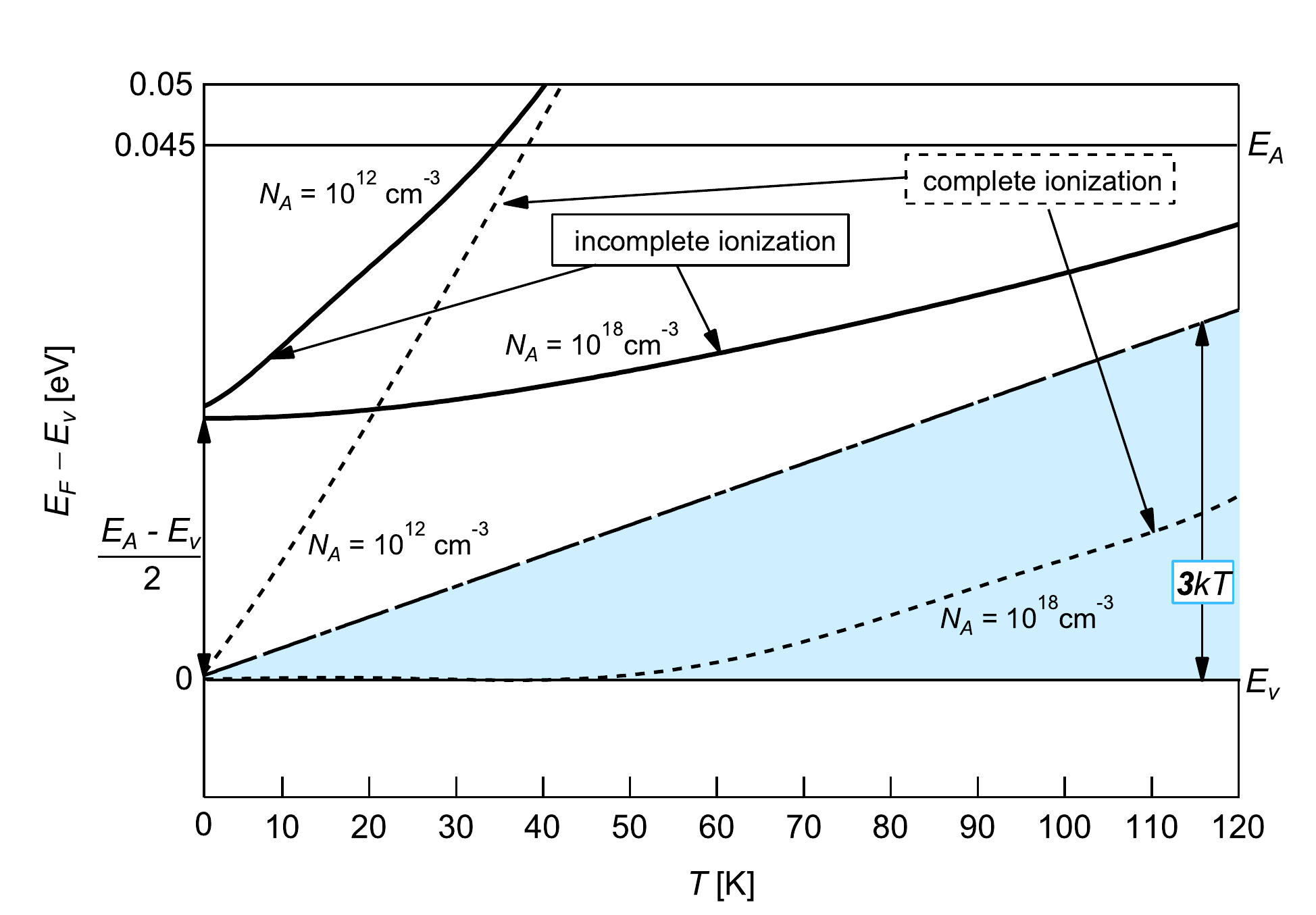}
	\caption{Fermi level position in the bandgap at cryogenic temperatures for silicon MOSFET with a $p$-type body. Validity of the Maxwell Boltzmann approximation only when incomplete ionization of the dopants in the MOSFET body is taken into account. Adapted from~\cite{tedpaper}. }
	\label{fig:fermi}
\end{figure*}
\section{MOSFET Modeling at Cryogenic Temperatures}
Industry-standard compact models are not readily available for deep cryogenic temperatures ($<$50\,K)~\cite{akturkdistributed}. While DC current characteristics down to 77\,K can still be adequately modeled, the embedded temperature-scaling in compact models is not sufficient to reach down to 4.2 K and lower temperatures using reasonable values for the physical parameters. This makes it clear that new physical phenomena popping up between 77 and 4.2\,K need to be investigated in MOSFET device physics first~\cite{kirschman,balestra,gutierrez2000low,balestra2001device}. Since the problem already arises in the DC characteristics, it is necessary to investigate the electrostatics and transport in MOSFETs at these extremely low temperatures~\cite{tedpaper}. A DC physical model that includes the missing phenomena certainly provides a more reliable basis for compact model development, especially when derivatives of the DC model will be required down the road for developing AC, RF, and noise models. In what follows, we first discuss the electrostatics and then the transport of MOSFETs at deep cryogenic temperatures. The first step in the derivation of the electrostatics is to check the validity of the Boltzmann relations of the mobile charge densities that are normally used in Poisson's equation. If the Maxwell-Boltzmann (MB) approximation of Fermi-Dirac statistics would not remain valid, complete Fermi-Dirac integrals would need to be used for the carrier distributions, complicating the development of an explicit model. Fortunately, for non-degenerate doping concentrations and deep-cryogenic temperatures, the MB approximation of Fermi-Dirac statistics remains valid~\cite{tedpaper}. As shown in Fig.\ref{fig:fermi}, the distance between $E_F$ and the band edges remains larger than $\approx 3k_BT$ in this temperature range (with $k_B$ the Boltzmann constant and $T$ temperature). We conclude that the Poisson-Boltzmann equation is a good starting point to derive the MOSFET electrostatics at deep-cryogenic temperatures. 

\begin{figure}[t!]
	\includegraphics[width=0.5\textwidth]{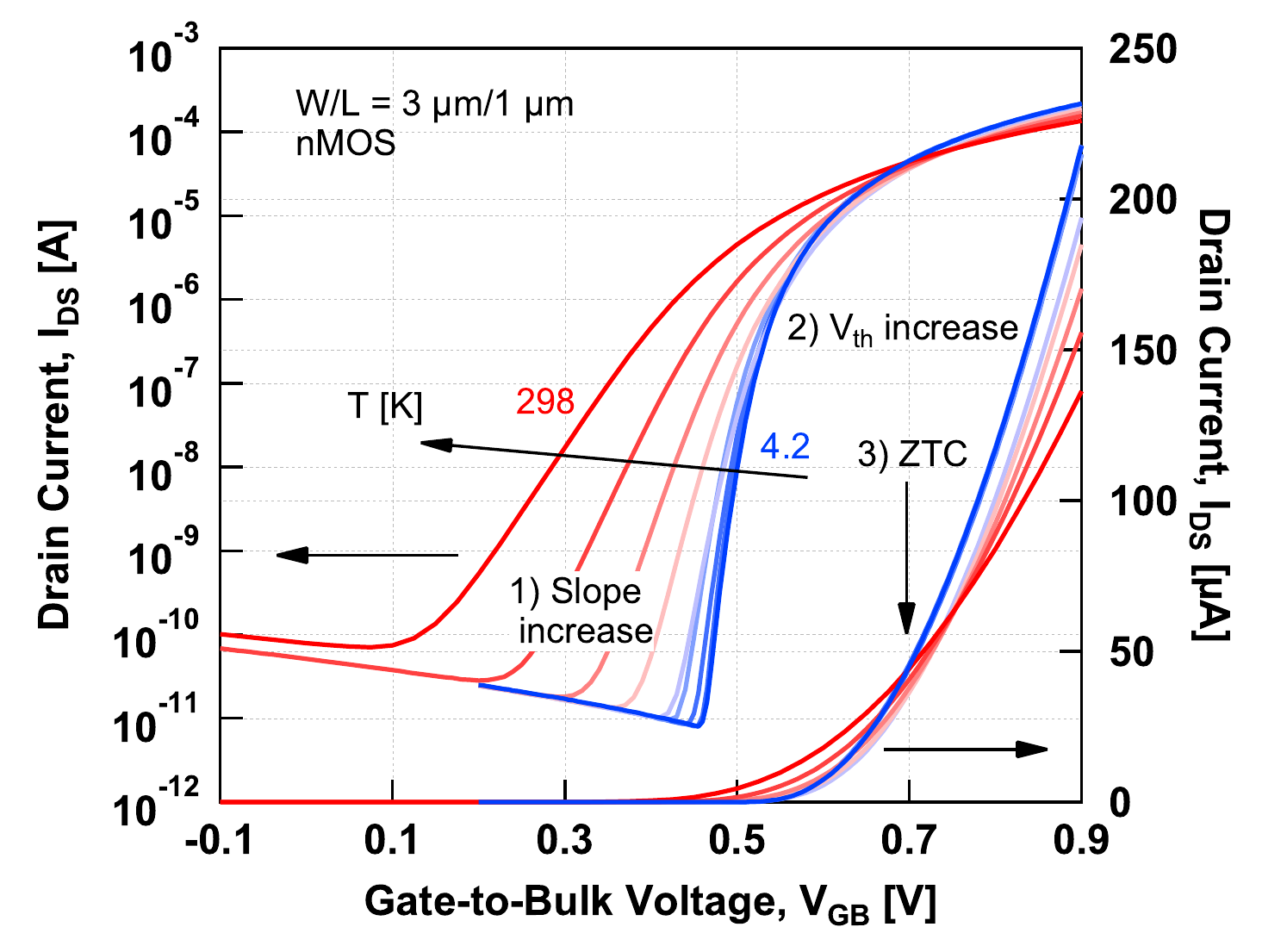}
	\caption{Low-temperature measurement results in a 28-nm bulk CMOS technology down to 4.2 K. Annotations highlight some typical observations: 1) improvement of the subthreshold slope, 2) increase in threshold voltage, 3)  zero temperature-coefficient (ZTC) bias point.
		Adapted from \cite{thresholdessderc}}
	\label{fig:meas}
\end{figure}

\begin{figure}[t!]
	\includegraphics[width=0.5\textwidth]{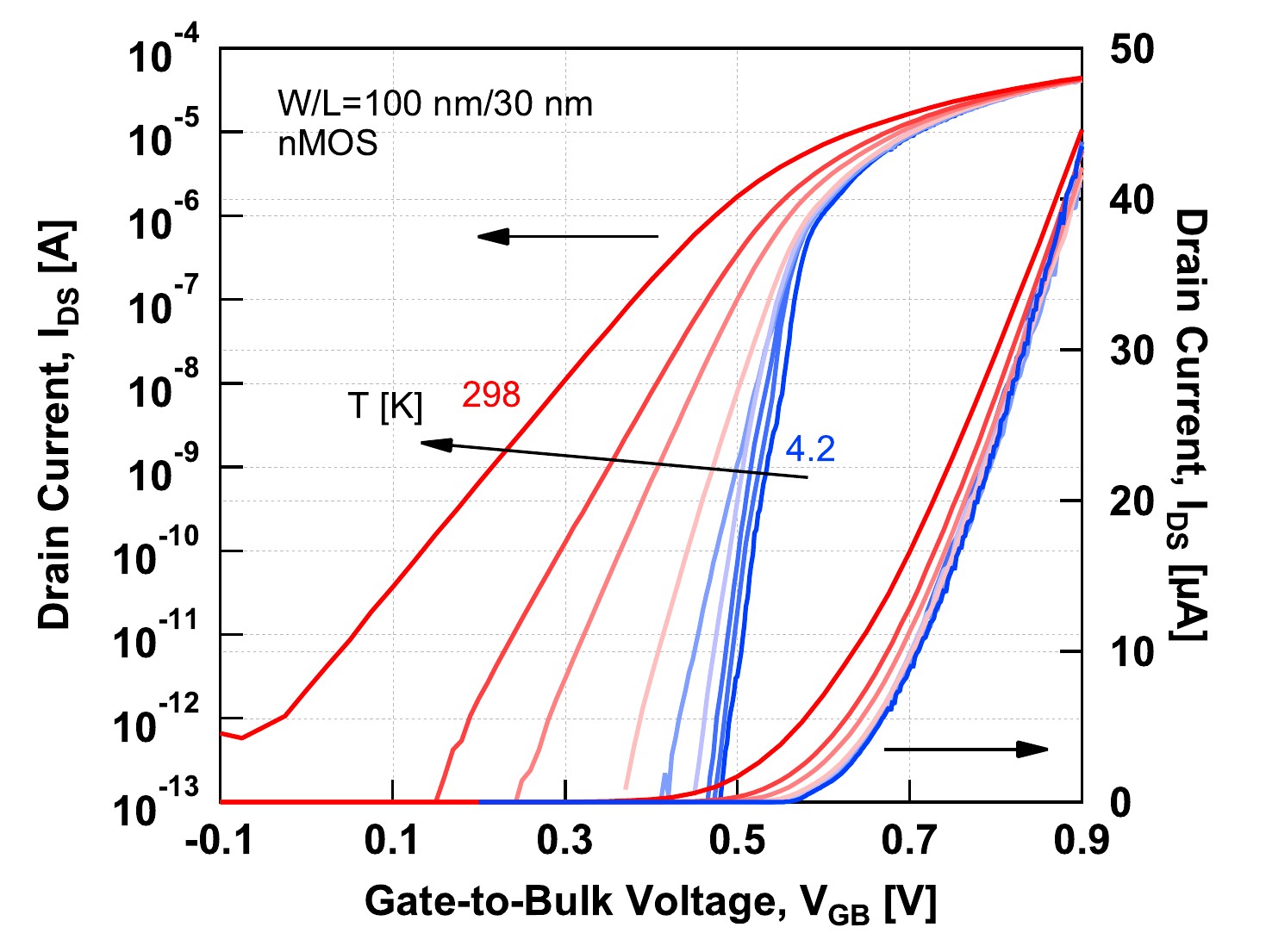}
	\vspace{-0.6cm}
	\caption{Low-temperature measurement results in a minimum-length device of a 28-nm bulk CMOS technology down to 4.2 K. Adapted from \cite{thresholdessderc}}
	\label{fig:meas2}
\end{figure}

\begin{figure*}[t!]
	\centering
	\includegraphics[width=0.7\textwidth]{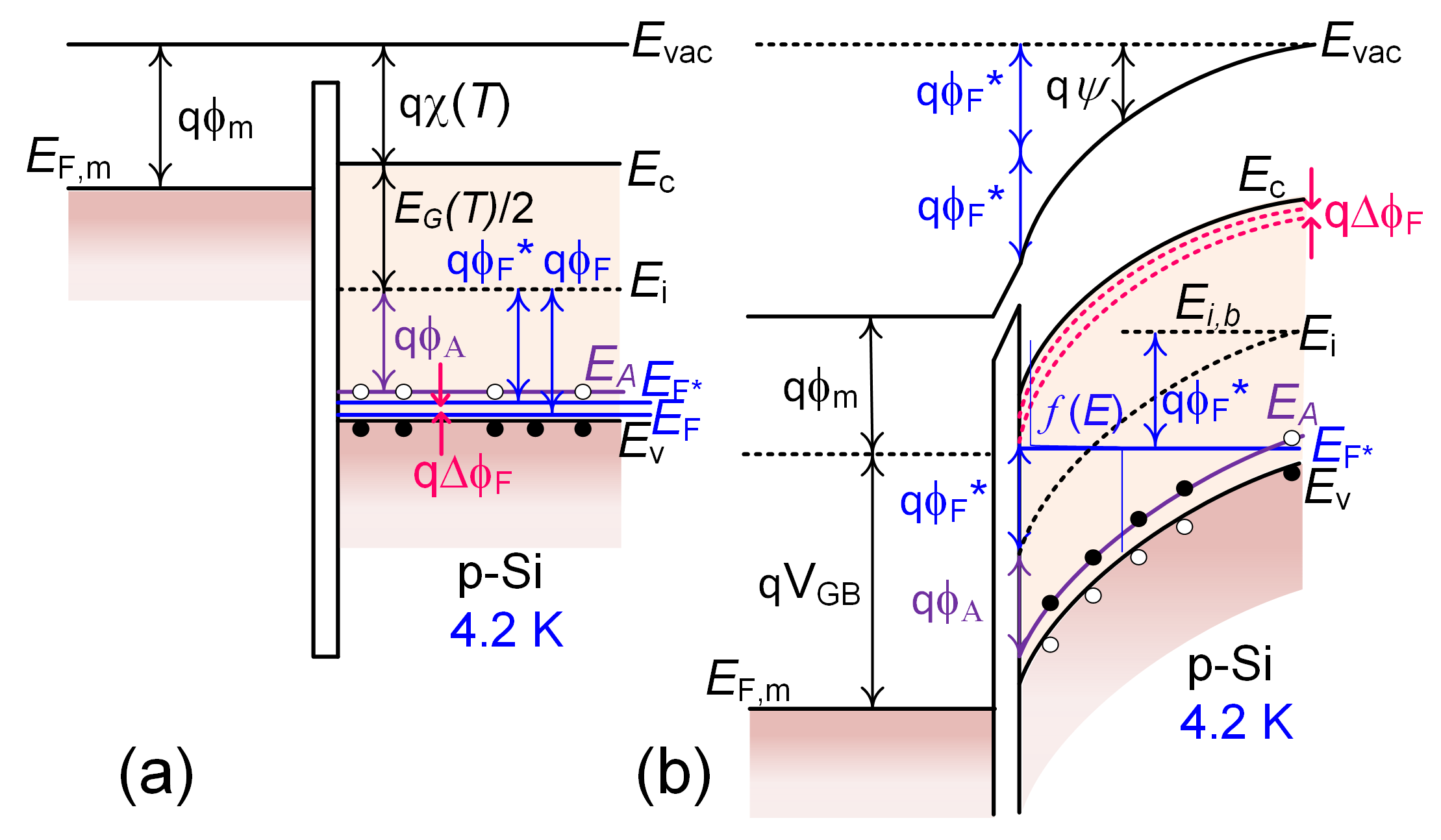}
	\caption{MOSFET band diagrams at 4.2\,K (a) flatband, (b) weak inversion. Adapted from \cite{thresholdessderc}.}
	\label{fig:banddiagram}
\end{figure*}

Dopants in the source and drain contacts can be assumed completely ionized at all times due to heavy doping effects~\cite{akturkinco,mott}. \textquoteleft Freezeout\textquoteright \, or thermal de-ionization of the dopants in the non-degenerately doped MOSFET body~\cite{substratefreezeout} can be included in the Poisson-Boltzmann equation simply by replacing $N_A$ with $N_A\times P(T,N_A)$ (where $N_A$ is the dopant concentration and $P(T,N_A)$ the ionization probability, considering an $n$-channel MOSFET). The ionization probability follows from semiconductor statistics: $P(T,N_A)$ is given by the Fermi-Dirac occupation probability $f(E)$ of the acceptor energy level $E_A$. Figure \ref{fig:fermi} highlighted the first consequence of incomplete ionization of the dopants in the MOSFET body: incomplete ionization ensures that the MB approximation remains valid ($| E_F$ $-$\,band edge$ | >\,\approx 3k_BT$)~\cite{tedpaper}. For complete ionization, $E_F$ would fall in the $3k_BT$ energy window below the band edge. The second consequence of freezeout (in the bulk) is that the inversion threshold is modified at each temperature and doping concentration from $2\Phi_\mathrm{F}$ to $\Phi_\mathrm{F}+\Phi_\mathrm{F}^*$, where $\Phi_\mathrm{F}^*<\Phi_{\mathrm{F}}$~\cite{thresholdessderc}. Therefore, although counterintuitive, incomplete ionization of the dopants \emph{lowers} the inversion threshold at each temperature below 298\,K. It is thus a misconception to assume that the increase in threshold voltage ($V_{th}$) in the $I-V$ characteristics for $T$ decreasing from room down to deep-cryogenic temperature (as in Figs.\,\ref{fig:meas} and \ref{fig:meas2}), is due to the initial dopant freezeout in the channel. In other words, the dopants that are thermally de-ionized in the channel in flatband and at deep-cryogenic temperature, say 4.2\,K, do not require more gate voltage to becomes ionized. The field-assisted ionization of the dopants is taken into account implicitly in the statistics of $f(E_A)$, but is already completed near flatband where the current is still below a measurable value at 4.2\,K. The increase in $V_{th}$ with decreasing $T$ is explained by the compound effect of exponential scaling of the Fermi-Dirac function $f(E)$ and the widening of the bandgap~\cite{odonnell_temperature_1991}, and thus not by the freezeout and/or field-assisted ionization of the dopants~\cite{thresholdessderc}. 
Besides dopants, another important element of electrostatics are the interface traps at the boundary between the channel and gate-oxide \cite{book1}. The importance of interface traps on the MOSFET's electrostatics at deep-cryogenic temperatures has since long been recognized~\cite{hafez_assessment_1990}. The density of interface traps in the bandgap typically increases toward the band edges~\cite{casse}. Since weak inversion at deep-cryogenic temperatures happens when the bands are bent such that $E_F$ at the interface is a few meV below the band edge (as shown in Fig.\,\ref{fig:banddiagram}), the number of interface states active in weak inversion can be substantially higher at deep-cryogenic temperatures than at room temperature. However, this increase should not be abused to explain the subthreshold-slope saturation. 

\textquoteleft Subthreshold-slope saturation\textquoteright \, means that the slope of the $I_D-V_G$ curve in subthreshold does not reach its maximum attainable value as theoretically predicted by the thermal limit for that temperature. The subthreshold-slope saturation has been measured in many types of FETs at deep-cryogenic temperatures~\cite{kamgar,shin, elewa,essderc,beckers_jeds,homulle,incandela,bohus_snw,designoriented,solidstate,galy,trevisoli_junctionless_2016}. It is usually shown by plotting the inverse subthreshold slope (or subthreshold swing $SS$) beside the Boltzmann thermal limit $SS=(k_BT/q)\ln10$. It is then clear that the trend of the measured $SS$ starts to roll off from the linear Boltzmann limit below $\approx$\,50\,K. This gives around 4.2\,K typically a $\Delta SS\approx$ 10\,mV/decade higher than the Boltzmann limit ($SS \approx$ 1\,mV/decade). This behavior continues down to temperatures as low as tens of millikelvin\cite{galy,incandela} Such high $SS$ can only be explained with the Boltzmann theory when anomalously high densities are used for the interface traps~\cite{galy}. This highlights that some physical phenomenon is lacking to explain $\Delta SS$. 

Band tails have been proposed to explain $\Delta SS$ by Bohuslavskyi \emph{et al.}~\cite{bohussat}. Due to disorder and finiteness of the crystal, band edges are not perfectly sharp. A density-of-states tail thus seeps into the bandgap, typically with exponential behavior~\cite{vanmieghem}. However, band-tail electrostatics and drift-diffusion transport is not sufficient to explain $\Delta SS$. Beckers \emph{et al.} have shown that the explanation for slope saturation does not lie in degraded electrostatics, but, rather, in an additional quantum transport mechanism which flows in parallel to the drift-diffusion current~\cite{jap}. This additional current component is a band-tail tunneling current (a hopping current through localized states in the channel) which becomes dominant over the drift and diffusion currents in weak inversion at deep-cryogenic temperatures. This hopping current has a worse subthreshold slope than the thermal limit from the diffusion current, and, thus, explains $\Delta SS$~\cite{jap}. Instead of $(k_BT/q)\ln10$, the limit of $SS$ at deep-cryogenic temperatures is derived as $SS=mW_t\ln10$, where $W_t$ is the characteristic extension of the exponential band tail in the bandgap (expressed in eV). $W_t$ is typically in the order of a couple of meV~\cite{jock}. The slope factor $m$ includes the interface-trap density. Using $SS=mW_t\ln10$ instead of $SS=m(k_BT/q)\ln10$ gives interface-state densities that do not reach anomalously high values at sub-Kelvin temperatures anymore~\cite{jap}. Furthermore, accounting for the hopping current leads to a $SS$ theory that explains the measured $SS$ roll-off from room down to sub-Kelvin temperature~\cite{jap}. 

\begin{figure}[b!]
	\centering
	\includegraphics[width=0.48\textwidth]{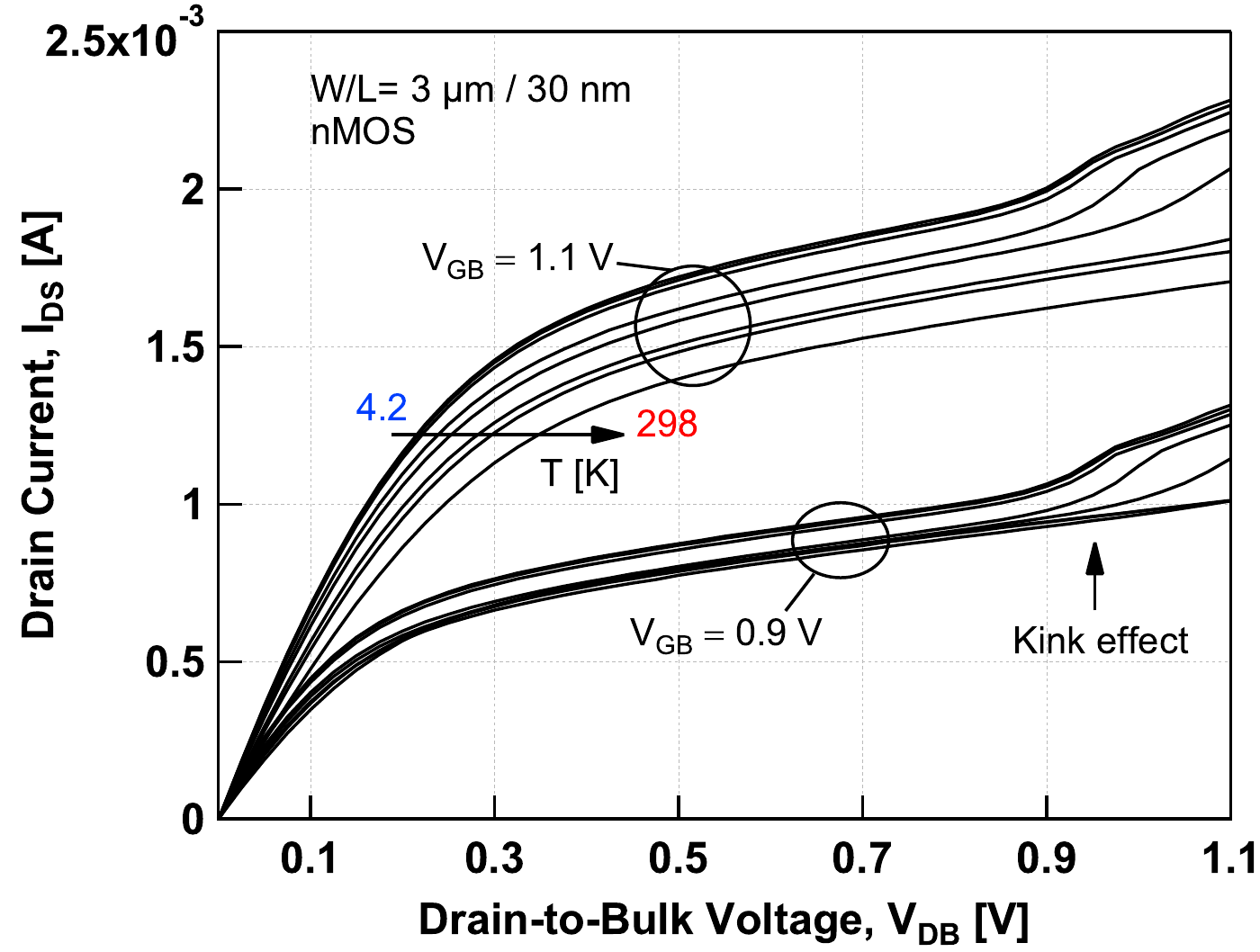}
	\caption{The kink effect sets in below $T=$\,110 K and above $V_{DB} = 0.9$\,V in the output characteristics of a commercial 28-nm bulk CMOS technology.}
	\label{fig:kink}
\end{figure}
Besides the saturation of the subthreshold slope, oscillations in weak inversion have been measured in the current characteristics of some devices fabricated in advanced and mature CMOS technologies\cite{simoen_parameter_1996,solidstate,bohus_snw}. These oscillations are most prominent at low drain bias (in the order of 1-10\,mV) and temperatures lower than $\approx$\,36\,K. Simoen \emph{et al.} hint at a resonant tunneling current to be responsible for this phenomenon~\cite{simoen_parameter_1996}. The oscillations are characterized by regions of negative-differential resistance, which is indeed a characteristic for resonant tunneling as exemplified by the well-known resonant-tunneling diode (RTD)~\cite{mizuta2006physics}. In an RTD, two potential barriers are typically created by stacking different materials~\cite{chang1974resonant}. For more tunnel barriers, this results in a \textquoteleft superlattice\textquoteright \, which can be exploited in the source of exploratory nanotransistors to obtain steep subthreshold slopes~\cite{gnani,jap_filtering}. In a standard commercial MOSFET, however, the manifestation of resonant tunneling is not immediately clear. Yet, a string of consecutive interface states in the channel can form a sequence of potential barriers between source and drain, and thus an accidental superlattice is established through which minority carriers can resonantly tunnel. Other defects than interface traps with active energies close to a band edge should also be considered for that purpose (e.g., dopants diffused unintentionally from the source and drain contacts into the channel~\cite{bohus_snw,zwanenburg2013silicon}).

Above, we have discussed the current components in weak inversion (drift, diffusion, hopping, and resonant tunneling). In strong inversion, the drift current dominates as usual. The main hurdle to overcome is then a mobility model \cite{Mob} that scales over temperature, gate and drain voltages, and device aspect ratios~\cite{emranione,emranitwo} The self-heating of the carriers is important as well in this bias regime~\cite{selfheating}. Furthermore, quantum confinement effects and ballisticity should be taken into account in an effective carrier mobility~\cite{colinge,baccarani, 8642438}. 

As a final note, the kink effect measured in the output characteristics at deep-cryogenic temperatures is not limited to mature CMOS technologies with micron length scales and high operating voltages. This phenomenon is also measured in contemporary commercial bulk CMOS technologies (e.g., 180-nm CMOS\cite{incandela}, and 28-nm CMOS). The appearance of the kink in 28-nm CMOS technology at deep-cryogenic temperatures is shown for the first time here in Fig.\,\ref{fig:kink}. The kink happens at an operating voltage above 0.9\,V. For 180-nm CMOS, the kink happens around 2\,V~\cite{incandela}. Since the performance is unstable around the kink, the cryo-CMOS designer should be aware of this phenomenon. The bias voltage should be limited depending on the technology. Phenomena that have not been part of our discussion are: mismatch~\cite{das,thart}, short-channel effects~\cite{plummer}, non-uniform doping~\cite{sze}, hysteresis~\cite{hysteresiskink}, hot-carrier degradation~\cite{woo} and other reliability phenomena.

\section{Conclusion}
In this review, we described different quantum computing implementations with a focus on the silicon initiative. This is mostly a brief review with short explanations of the underlying physics for different aspects of quantum computing. The importance of developing low-temperature peripheral electronics and understanding the solid state physics in MOSFETs at deep cryogenic temperatures were explained. Even though the field of quantum computing and the infrastructure around it is still in its infancy, quantum computation, teleportation, and cryptography are nowadays already commercially available.

\newpage
\bibliography{bare_conf}
\bibliographystyle{ieeetran}
\newpage

\end{document}